\documentstyle[preprint,aps]{revtex}
\newcommand{\be}{\begin{eqnarray}}
\newcommand{\ee}{\end{eqnarray}}

\draft
\begin{document}
\def\thefootnote{\fnsymbol{footnote}}
\preprint{\begin{minipage}{5cm}
\flushright hep-ph@xxx/9708405\\
UAHEP9713\\ TUFTS TH-97-G02\\
\end{minipage}}
\vspace*{20mm}
\title{Top Quark Analysis in the Light Gluino Scenario}
\author{L. Clavelli\footnote{e-mail: lclavell@ua1vm.ua.edu}
and Gary R. Goldstein\footnote{e-mail: goldstei@pearl.tufts.edu}}
\address{
$*$ Department of Physics and Astronomy\\ University of Alabama\\
Tuscaloosa, Alabama 35487 \\}
\address{
$\dagger$ Department of Physics and Astronomy\\
Tufts University\\
Medford, MA 02155\\}
\date{August 1997}
\maketitle
\noindent
\begin{abstract}
The Fermilab top quark analysis is heavily dependent on the assumption
of standard model backgrounds only. In the light gluino scenario, the
stop quarks lie near the top in mass and their decays can influence the
resulting top quark mass by an amount that is not small relative to the
currently quoted errors. Several slight anomalies in the top quark
analysis find a natural explanation in the light gluino case.\\
\thispagestyle{empty}
\setcounter{page}{0}
\end{abstract}
\narrowtext
\par
In the past few years the Tevatron at Fermilab has provided convincing
evidence for physics beyond the topless standard model
\cite{FNAL}.
The events contain isolated leptons, missing energy, and evidence for b
quark jets all of which are part of the expected signal for top quark
production and decay. Analyzed within the context of the standard model
the best fit to a top quark mass and production cross section gives
\be
m_t = 175.6 GeV \pm 5.7 GeV (stat) \pm 5.9 GeV (sys) \label{topmass}
\ee
\be
  \sigma(t {\overline t},(m_t=175 GeV)) = 7.5 ^{+1.9}_{-1.6} \quad pb
  \quad .
\label{prodcs}
\ee
\par
Due to the dependence of experimental acceptances and efficiencies on
the top quark mass, the experimental production cross section rises with
decreasing top mass, becoming $10.0 \pm 1.4 pb$ for a top quark mass of
$160 \quad GeV$. Furthermore, it is clear that these values are strongly
dependent on the assumption that the background is correctly given by
the standard model which predicts very few events of the type seen in
the mass region near 175 GeV. Indeed, in the absence of a specific model
beyond the standard, no other assumption is possible. For this reason,
if for no other, it is useful to construct a specific testable
alternative to the standard model. The observation of b quarks in the
events, however, severely constrains non-standard interpretations. The
top sample is still a low statistics set and significant fluctuations
are to be expected. Nevertheless, several slightly unsettling features
of the sample have been noticed \cite{DG,Sliwa} which could be taken as
hinting at
effects beyond the standard model or, at least, as indicating
in which directions there is room for such effects. Among these are the
facts that\\
\par
1) The joint probability that all the "top events" are due to a single
quark of any given mass seems small in any of the standard
monte-carlos. Put in another way, the spread in apparent top quark mass
on an event by event basis is much greater than in the standard model
monte-carlos with a $175 GeV$ top quark \cite {DG}.\\
\par
2) There seems to be a systematic tendency for the events in which both
tops decay leptonically to suggest a smaller mass than those in which a
single top decays leptonically which again suggest a smaller top mass
than the non-leptonic decay events. One finds \cite{DG,Lys} from the
di-lepton events $m_t=(162 \pm 21 \pm 7) GeV$, from the single lepton
events $m_t =(176 \pm 4.4 \pm 4.8) GeV$, and from the hadronic events
$m_t = (187 \pm 8 \pm 12) GeV$. \\
\par
3) The CDF and D0 results for the top production cross section seem to
be somewhat higher than the best theoretical estimates \cite{Catani} for
a 175 GeV top quark:\\
\be
 \sigma_{th} (m) = ({\exp{(175-m) \over 31.5}})\quad (4.75 ^{+.73}_
 {-.62} pb)
\label{sigma}
\ee
to be compared with the experimental result (primarily from the single
lepton events) given in eq. \ref{prodcs}.\\
\par
4) In some of the events, the invariant mass of two jets identified as a
non-leptonic W decay do not well reproduce the W mass.
\cite{Giromini}.\\
\par
5) The $t {\overline t}$ system seems to be produced with somewhat
greater transverse energy and to be accompanied by more extra jet
activity than expected in the standard model \cite{Sliwa}.\\
\par
While emphasizing again that all of these "effects" could easily
disappear with better statistics, it is interesting to consider the
effect on the top analysis of specific models for supersymmetry (SUSY)
in the top region. Standard SUSY scenarios, such as that of squarks and
gluinos in the $330$ GeV region can, at best, \cite{BarnettHall} account
for one or two of the anomalous top events. Recently, however,
phenomenological hints have been noted suggesting squarks in the
region below $200$ GeV.
\cite{Kaneetal,CT}.  The effect of these scenarios on
the top analysis needs to be considered in detail.
\par
In this article, we present the predictions of the light gluino (LG)
scenario in which both the universal gaugino mass $m_{1/2}$ and the
trilinear coupling parameter $A$ are set to zero. This is a
configuration of special symmetry in the supergravity (SUGRA) related
SUSY breaking model and results in gluino and photino masses below $1$
GeV. The remaining parameters will be constrained by
phenomenological and theoretical requirements below.
Problems for the LG scenario are posed by the $\tau$ decay data
\cite{tau} and by the ALEPH four-jet angular distribution \cite{fourjet}
data. These analyses are, however, each vulnerable to criticism
\cite{farrar97} and many
other phenomenological observations have been noted as supporting the LG
idea \cite{many}.
\par
With $m_{1/2}$ and $A$ set to zero, the SUGRA-related SUSY standard
model has four main parameters: a universal scalar mass $m_0$, the Higgs
mixing parameter $\mu$, the ratio $\tan\beta$ of the Higgs vacuum
expectation values, and the top quark mass $m_t$. Two of these
parameters are tightly constrained in the LG scenario by the
phenomenological requirement that the chargino and neutralino masses be
above half the Z mass.
\be
 M_{\tilde \chi^{\pm}}^2 = M_W^2+\mu^2/2 \pm \sqrt{({\mu^2 \over 2}+M_W^2)^2
     - M_W^4 \sin^2(2 \beta)} \quad > \quad M_Z^2/4
\label{chargino}
\ee
\be
   M_n = {{2 \sqrt{3}}\over\sqrt{M_Z^2+mu^2}} |\cos((\phi+ 2 \pi n)/3)|
\ee
\be
   \cos(\phi) = - {b \over 2} (3/a)^{3/2}
\ee
with
\be
    b = \mu \sin(2 \beta) /M_Z \\
    a = 1 + \mu^2/M_Z^2  \quad .
\ee
\par
The result is that $|\mu|$ is constrained to be below $M_W$ and
$\tan\beta$ cannot be far from $1.6$. It is interesting to note that
this value of $\tan\beta$ is in one of two ranges preferred by grand
unification studies. Similarly, without relying on the theoretical
guidance provided by the SUGRA model, \cite{Kaneetal} arrive at similar
values of $\mu$ and $\tan\beta$ on purely phenomenological grounds.
The neutralino spectrum, apart from an ultra-light photino, is predicted
in the LG scenario to begin in the $50$ GeV region
\cite{ClavelliMPL} very close to the
values postulated on phenomenological grounds in Ref.\cite{Kaneetal}.
In this latter approach, non-universal (and ad hoc) gaugino
masses, $m_1$, $m_2$, and $m_3$, are introduced while keeping the gluino
heavy.
\par
A second strong constraint in the LG scenario comes from the attractive
assumption of radiative breaking of the electroweak symmetry. In this
picture, which in fact is difficult to avoid, one of the Higgs squared
masses runs to negative values thus triggering the electroweak breakdown
near the SUSY scale. The required value of $\mu$, which we label here as
$\mu_{rad}$ is in lowest order related to the other parameters of the
theory by
\be
 \mu_{rad}^2 = -m_Z^{2}/2 - m_0^2 - m_0^2
 {{({m_t \over{205 GeV}})^2 \over{\cos 2\beta}}} {f \over 2}
\label{radbreaking}
\ee
where
\be
  f = 3 + {({A \over {m_0}} + {\mu_{rad} \over{m_0 \tan \beta}})^2}
  (1 - {({m_{t} \over{205 GeV}})^2 \over \sin ^2(\beta)}) \quad .
\ee
Since the third term of eq. \ref{radbreaking} must overcome the first two
negative terms to equal the positive definite left hand side, if $\mu^2$
is small as required by eq. \ref{chargino}, the radiative breaking can be
satisfied only for highly constrained values of $m_t$. This relation
requires unacceptable fine-tuning if $m_0$ is large. We seek solutions
for $95 GeV < m_0 < 150 GeV$ and $130 GeV < m_t < 180 GeV$. The observed
dijet angular distributions at Fermilab rule out squarks in the LG
scenario with masses between $150$ and $650 GeV$ \cite{dijets}. Values
of $m_0$ below $95$ are most likely inconsistent (in the LG case) with
measurements at LEP-2.
\par
Since this is a perturbative result, we require only that $\mu_{rad}$ be
equal to the $\mu$ of eq. \ref{chargino} to within $10\%$. The stringent
results of \cite{Lopez} are then relaxed. Still we find that the
constraints can only be satisfied for $m_t < 169 GeV$ and $m_0<142 GeV$
and, if $m_t > 150 GeV$, then $m_0 < 120 GeV$. The low top masses found
by generating events in the space of
parameters via a Monte-Carlo scheme
coincide with the apparent top mass seen experimentally in the
di-lepton decays. We would therefore like to investigate whether the
higher top masses seen in the single lepton and hadronic channels are
due to contamination from top squark decays. The low output values of
$m_0$ and the resulting low squark masses coincide with those suggested
by the low jet $E_T$ and scaling anomalies seen at Fermilab \cite{CT}.
Such low squark masses are excluded in the heavy gluino case by direct
searches for the expected decays into isolated energetic leptons and
missing transverse energy
\cite{FNALlimits}.  In the LG scenario, however, the squarks will
decay primarily into a quark-gluino dijet thus evading the direct
searches in the lepton plus missing energy channel.
\par
In the SUGRA model the up type squark mass matrices are given by
\be
   M_{\tilde q_L}^2 = M_0^2 + M_q^2 + M_Z^2 \cos(2 \beta)
   \bigl({1 \over 2} - {2 \over 3} \sin^2(\theta_W) \bigr)
\label{squarkLmass}
\ee
\be
 M_{\tilde q_R}^2 = M_0^2 + M_q^2 + {2 \over 3}M_Z^2 \cos(2 \beta)
 {\sin^2(\theta_W)} \quad   .
\label{squarkRmass}
\ee

For each flavor there is also an off-diagonal term

\be
  M_{LR}^2 = M_q (A + {\mu \over \tan(\beta)}) \quad   .
\label{LRmass}
\ee
\par
In \cite{ClavelliMPL} non-zero values of A were considered with the
result that the lightest stop quark could be made significantly lighter
than the top quark due to the off-diagonal term in the mass matrix. This
would then allow a large stop quark related enhancement in the $Z$ decay
into $b$ quarks. In the heavy gluino case of the constrained SUSY model
the large off-diagonal term would not by itself give light stop quarks
due to a large diagonal contribution proportional to $m_{1/2}^2$. Now
that the $R_b$ anomaly has largely disappeared we can consider the case
of zero A which is much more natural in the LG scenario. The
off-diagonal term is then given by the $\mu$ parameter with the result
that the stop quarks can be predicted to be both near or above the top.
\par
As a final constraint on the parameter space we consider the electroweak
$\rho$ parameter which measures the relative strength of neutral to
charged currents. $\rho$ differs from unity in the presence of
non-degenerate weak doublets. The large $t-b$ splitting makes the $\rho$
parameter sensitive to the top quark mass. The current experimental
value of $\rho$ is essentially saturated by a $175 GeV$ top quark
leaving little room for SUSY contributions \cite{Rosner}.
\be
  \delta \rho = .0095 \cdot({m_t \over {175 GeV}})^2
                 + \delta \rho_{SUSY} = .0095 \pm .0014 \quad   .
\label{rhoparam}
\ee
This suggests that the squarks and sleptons are very degenerate as
suggested by the SUGRA inspired mass matrices as given in
eqs. \ref{squarkLmass}, \ref{squarkRmass}, \ref{LRmass}. If one abandons
the SUGRA universality conditions, a $\rho$ parameter near the standard
model value could become unexplained unless the SUSY particles are much
higher in mass. In fact, imposing the $\rho$ parameter constraint
already further constrains the parameters of the SUSY model for values
of $m_0$ in the range we are considering. As can be seen from
eq. \ref{rhoparam} any non-degeneracy of the squarks and sleptons tends to
reduce the top mass if the $\rho$ parameter constraint is to be
preserved. Using the one-loop SUSY contributions from \cite{Djouadi} and
imposing the experimental constraint of eq. \ref{rhoparam} the solution
space is further restricted to
\be
     m_t < 162 GeV\\
     m_0 < 133 GeV  \quad   .
\ee
\par
A top mass or scalar mass outside of this range would require abandoning
the light gluino scenario or relaxing at least one of the other
assumptions discussed above such as the radiative breaking constraint or
the universality of scalar and gaugino masses. We note that those
studying the heavy gluino case have already been lead to abandon the
SUSY breaking mass universality relations while, in the light gluino
case, it is still possible and interesting to maintain them. In addition
we suspect that, in the heavy gluino scenario of \cite{Kaneetal} with
stop and sneutrino in the $50 GeV$ region and other squarks in the
$200-300 GeV$ region, the $\rho$ parameter constraint will also force
the top quark to low values such as those above. On the other hand, in
this heavy gluino case, it is not clear whether such a lower top quark
mass can be made consistent with the Fermilab data.
\par
With the parameter space now tightly constrained we would like to
discuss the phenomenology of the top quark region. The question now
becomes what are the stop quark masses and what are their decay chains?
In the current model the stop quarks are each almost equal mixtures of
left and right handed stops with the lighter stop quark being 0.3 to 33
GeV above the top and the heavier stop being in the range $183 GeV
<m_{\tilde t_2}<209 GeV$. These predicted stop quark masses are too high
to cause a large $b$ excess in $Z$ decay. Nevertheless,
in the current model, there will be a (largely) flavor-independent
enhancement of the hadronic decay rate of the $Z$ due to virtual
squark-gluino corrections which will make the apparent value of
$\alpha_s$ measured at the $Z$ higher than the actual one.  The
predicted hierarchy of masses and dominant decays are shown in table 1.
\vskip 0.5 in
\halign{ # & \quad # \quad & # \cr
Particle & Mass & prominent decay modes \cr photino & $< 1 GeV$ & stable
or $ goldstino + \gamma$ \cr gluino & $< 1 GeV$ & ${\tilde \gamma} +
hadrons$ \cr & & possibly $goldstino + gluon$ \cr chargino & $\simeq 50
GeV$ & $q {\overline q} {\tilde g}$ \cr neutralino& $\simeq 50 GeV$ & $q
{\overline q} {\tilde g}$ \cr u,d,s,c,b squarks & $\simeq 110 GeV$ & $q
{\tilde g}$ \cr sleptons & $\simeq 110 GeV$ & $\ell {\tilde \gamma} ,
\ell \chi$\cr top quark & $<162 GeV$ & $W b$ \cr ${\tilde t_1}$ & $m_t
+(.3 \sim 33 GeV)$ & $t {\tilde g},b{\tilde \chi}$ \cr ${\tilde t_2}$ &
$183 \sim 209 GeV$ & $t {\tilde g} , b {\tilde \chi}$\cr }
\vskip 0.2in
\noindent
Table 1. mass and decay channnels for SUSY particles in the light
gluino scenario ($m_{1/2}=A=0$) assuming radiative breaking and $\rho$
parameter constraints.
\vskip 0.5in
\par
The lightest chargino and neutralino in this model as well as the
squarks near $110 GeV$ have predominantly hadronic decay modes with only
rare decays into leptons plus missing energy thus evading previous SUSY
searches. The heavier chargino will have a prominent decay into $W
{\tilde \gamma}$. Thus pair production of this chargino or of the
sleptons could lead to
events of the form $\ell ^{+} \ell ^{-} \gamma \gamma + invisible$
discussed by \cite{Kaneetal} providing the ${\tilde \gamma}$
decays into $\gamma$ + Goldstino or gravitino as treated for example
in \cite{Yuanetal}. The top quark decays predominantly in the standard
model mode $W+b$. The possible decay into ${\tilde c}+{\tilde g}$ is
presumably highly suppressed by Kobayashi-Maskawa angles. The stop
quarks decay predominantly into $t+{\tilde g}$ and $b+{\tilde \chi}$ with
approximate relative branching ratios
\be
B_i = {{\Gamma ({\tilde t_i} \rightarrow t + {\tilde g})}\over {\Gamma
({\tilde t_i} \rightarrow b + {\tilde \chi})}} = 4 {\alpha_s \over \alpha}
\sin^2 \theta_W {{(M^2_{t_i}-M^2_t)^2} \over {(M^2_{t_i}-M^2_{\tilde
W})^2}} { 1 \over {1 + m_{t}^2/m_{W}^2}} \quad .
\ee
We have included the final factor here as an estimate of the effect of
Higgsino admixture in the chargino since the coupling of this component
is proportional to the top mass. The corresponding leptonic branching
ratios of the stops are
\be
    B_{L,i} = {{B_{i}B_{L,t}} \over {1+B_i}} \quad  .
\ee
Here $B_{L,t} \simeq .24$ is the inclusive (prompt) electron plus muon
decay branching ratio of the top including feed-down from $\tau$. Due to
phase space, the stops, especially the lighter, decay preferentially
into $b+chargino$ which, according to table 1, very rarely leads to a
high energy lepton plus missing energy. Thus the stop production will
lead primarily to non-leptonic, b-containing events. The total invariant
mass on each "side" will be above the top mass but if the lowest energy
jet from the chargino is partially or completely discarded due to
experimental cuts, the apparent "top" mass on an event by event basis
could vary over a large range as, in fact, seems to be the case with the
Fermilab top events. If the full jet energy is collected it should be
possible to observe a a peak at the chargino mass ($\sim 50 GeV$) instead of
at the $W$ mass. It has been suggested \cite{Farrar} that this $50 GeV$
chargino is responsible for the anomalous events in the Aleph four-jet
sample.
\par
Since both the $t$+gluino and the $b$+chargino decays of the stop have
an extra (possibly low energy) jet relative to the standard model $t
{\overline t}$ production, one can expect the top quarks reconstructed
in a standard model analysis to be accompanied by extra jet activity and
to exhibit a greater than expected transverse momentum.
\par
Furthermore, since the $B_{L,i}$ are small and the stop masses are above
the top mass, the di-lepton decays are almost always attributable to
direct $t {\overline t}$ production with only a small contribution from
stop initiated events. The average reconstructed mass in the di-lepton
events should then be only slightly higher than the true top quark mass.
The single lepton events will have a somewhat higher contribution from
stop pair production with one stop decaying non-leptonically and the
other decaying to a top with a subsequent leptonic decay. Since the mass
attributed to the top in such events comes from the hadronic side, the
effective top mass in single-lepton events will tend to be higher than
in the di-lepton events. The gluon appearing in the stop to top decay
could, in at least some fraction of the events, be mis-associated with
the non-leptonic decay of the assumed recoiling top thus leading to a
further enhancement of the apparent top mass in single lepton events.
\par
To properly model these effects it would be necessary to construct a
hadronization monte-carlo with the mass and coupling information of the
light gluino model and including the effects of detector acceptances.
With light gluinos and stops in the top region, there are important SUSY
corrections to the top production cross section Since such a monte-carlo
is not available at present and since, in any case, the top production
cross section would depart from the SM predictions due to light gluino
and stop loops, we make the following preliminary model to estimate the
possible size of expected effects. For each value of the four parameters
$m_0, \mu ,\tan(\beta)$, and $m_t$ and the consequent masses $m_{t_1},
m_{t_2}, M_{\tilde \chi}$ we define three pair production cross sections
\be
     \sigma_t = \sigma_{th}(m_t) \\
     \sigma_{t_1} = \sigma_{th}(m_{\tilde t_1})/2 \\
     \sigma_{t_2} = \sigma_{th}(m_{\tilde t_2})
\label{eq:stopprods}
\ee
where $\sigma_{th}(m)$ is the empirical fit to the theoretical top
production cross section for mass $m$ given in eq. \ref{sigma}. We enhance
the heavier stop production (by an estimated factor of $2$)
due to more important contributions from processes such as
\be
     q {\overline q} \rightarrow b {\tilde \chi} {\tilde t} \\
     q {\overline q} \rightarrow t {\tilde g} {\tilde t} \quad  .
\ee
With light gluinos and charginos, these processes have a mass advantage
over the stop pair production which is not shared to the same extent by
the analogous top or lighter stop production processes. A further
enhancement of stop production above the crude estimate of
eq. \ref{eq:stopprods} would increase
the difference between the apparent "top" masses seen in the different
decay topologies. In subsequent studies of the scenario outlined here it
will be important to incorporate a more precise calculation of the stop
production cross sections and decay branching ratios as functions of the
masses in the theory. Nevertheless, we content ourselves at present with
the simple model discussed here in order to illustrate the approximate
size of the expected effect.
\par
We assume that each top initiated event will produce a reconstructed top
mass $m_t$, and each $t_1$ or $t_2$ initiated event will produce a
reconstructed mass $t_1$ or $t_2$. This is clearly oversimplified but is
motivated by the expectation that, in the single lepton events, the mass
is reconstructed from the hadronic side and that the extra jet activity
from the stop decays will increase the measured top mass even in the
case of stops to tops to dileptons. As it turns out, however the stop
contribution to the di-lepton events is very small. The di-lepton
events, single lepton events, and non-leptonic events will be produced
with cross sections
\be
 \hfil \sigma_{2L}&=& B_{L,t}^2 \sigma_{t} + B_{L,1}^2 \sigma_{t_1} +
       B_{L,2}^2 \sigma_{t_2}\hfil\\
 \hfil\sigma_{1L}&=& 2 \Bigl( B_{L,t} (1-B_{L,t}) \sigma_{t} +
  B_{L,1} (1-B_{L,1}) \sigma_{t_1}+  B_{L,2} (1-B_{L,2}) \sigma_{t_2}
     \Bigr)\hfil\\
 \hfil\sigma_{had}&=& (1-B_{L,t})^2 \sigma_{t} + (1-B_{L,1})^2 \sigma_{t_1}
         + (1-B_{L,2})^2 \sigma_{t_2} \quad .\hfil
\ee
The reconstructed masses in the di-lepton, single lepton, and hadronic
events, will then be roughly:
\be
  M_{2L} &=& \Bigl( m_t B_{L,t}^2 \sigma_{t} + m_{t_1} B_{L,1}^2
  \sigma_{t_1} + m_{t_2} B_{L,2}^2 \sigma_{t_2} \Bigr)/\sigma_{2L}\\
  M_{1L} &=& 2 \Bigl( B_{L,t} (1-B_{L,t}) m_t \sigma_{t}  + B_{L,1}
  (1-B_{L,1}) m_{t_1} \sigma_{t_1} + B_{L,2} (1-B_{L,2}) m_{t_2}
  \sigma_{t_2}\Bigr)/\sigma_{1L}\\
  M_{had} &=& \Bigl( (1-B_{L,t})^2 m_t \sigma_{t} + (1-B_{L,1})^2 m_{t_1}
  \sigma_{t_1} + (1-B_{L,2})^2 m_{t_2} \sigma_{t_2}\Bigr)/\sigma_{had}
  \quad .
\ee
Since the Fermilab top measurements are dominated by the single lepton
events, the reported top production cross section is approximately
\be
   \sigma = {\sigma_{1L} \over{2 B_{L,t} (1-B_{L,t})}} \quad   .
\label{sigma1L}
\ee
We make, therefore, a final loose cut in our monte-carlo requiring that
$4 pb < \sigma <12 pb$ and that $\sigma_{had}/(1-B_{L,t})^2 < 20 pb$.
In the simplified model presented
here we find $9.3pb < \sigma < 12 pb$ (upper limit imposed)
which is consistent with the experimental values for a $160 GeV$ top.
The effective cross section
in the di-lepton channel is about $1.5 pb$ lower and the effective
cross section in the hadronic channel is greater than $16 pb$.
The apparent values of the experimental top production cross section
as measured in the three channels are given in \cite{Gerdes}.
Although these experimental cross
sections are consistent within errors with the standard model they are
also consistent with a larger apparent top cross section in the fully
hadronic channel as predicted here in the simplified model.
\par
Although we have relied here on crude estimates of stop production cross
sections and branching ratios, in
the $m_{1/2}=A=0$ version of the supergravity-inspired SUSY breaking
model which leads to light gluinos and photinos, when experimental
constraints from LEP and Fermilab are imposed, the following conclusions
can be drawn.
\par
1) The top quark and stop quark masses are
\be
     m_t = 157 \pm 4 GeV \\
     .3 GeV < m_{t_1}-m_t < 21 GeV \\
     198 GeV < m_{t_2} < 207 GeV
\ee
\par
2) "Top-like" events at Fermilab will exhibit a range of apparent masses
of the top quark with the average masses of the three topologies
satisfying $M_{2L} < M_{1L} < M_{had}$. In the simple model presented
here:\\
\be
     M_{2L} \simeq 161 \pm 4 GeV \\
     M_{1L}-M_{2L} \simeq 5 GeV\\
     M_{had}-M_{1L} \simeq 5 GeV
\ee
\par
3) Some of the "top" events will be associated with extra jet activity
and total transverse momentum above that expected in the standard model.
\par
4) Some of the events attributed to a non-leptonic top decay will not
well reconstruct the $W$ mass. In some of these events the apparent $W$
mass will be larger than expected due to contamination from the
predicted extra low energy jets. However, it should also be possible to
find evidence for a $50 GeV$ chargino recoiling against a $b$ quark jet.
\par
5) In the most likely scenario where SUSY decays of the top quark are
highly suppressed, all of the "top" events will have $b$ quark jets in
their decays.
\par
6) Apart from the stop quarks, squarks and sleptons will be in the $100$
GeV region but will not have prominent decays into isolated leptons plus
missing energy. This coincides with the postulated mass of the charged
sleptons in \cite{Kaneetal} although, for phenomenological reasons in
the heavy gluino case, these authors postulate a significantly lower
sneutrino mass and a significantly higher squark mass.
\par
7) Apart from the gluino and photino which will be in the ultra-low mass
window, the tree-level gaugino masses are expected to be
\be
    46 GeV < m_{\chi_1^{\pm}} < 51 GeV\\
    116 GeV < m_{\chi_2^{\pm}} < 131 GeV\\
    46 GeV < m_{N_1} < 51 GeV\\
    76 GeV < m_{N_2} < 87 GeV\\
    122 GeV < m_{N_3} < 137 GeV \quad .
\ee
These masses as well as the accompanying mixing angles are predicted
before imposing constraints from the Fermilab "top quark" measurements.
The squark and slepton masses are determined (after loosely imposing the
Fermilab constraints) by the output parameters:
\be
   95 GeV < m_0 < 118 GeV\\
   1.50 < \tan(\beta) < 1.69\\
   39 GeV < | \mu | < 79 GeV \quad .
\ee
\vskip 0.5in
\par
The main purpose of the present paper has been to set forth the
sparticle mass predictions of the $m_{1/2}=A=0$ model subject to
constraints from radiative breaking, $\rho$ parameter measurements, and
cross section measurements at LEP and Fermilab. We have noted that the
predicted mass hierarchy with, in particular, stop quarks in the $160$
to $210 GeV$ region shows promise for explaining several possible
anomalies in the top quark region. The model predicts a top quark mass
several standard deviations below the $175 GeV$ mass resulting from a
standard model analysis of the Fermilab events. A complete analysis of
the scenario suggested in this paper is beyond the scope of the current
work and will require the incorporation of the light gluino effects into
a full hadronization monte-carlo with detailed treatment of experimental
cuts. We feel that the light gluino scenario provides testable
predictions for the top quark region that, at a minimum, are not ruled
out by current measurements.
\par
In the course of this analysis we profited from discussions with K.
Sliwa of Tufts University. This work was supported in part by the
Department of Energy under grant $DE-FG02-96ER40967$
at the University of Alabama and $DE-FG02-92ER40702$ at Tufts
University.
LC would like to thank the Department of Physics at Tufts for
hospitality during the summer of 1997 when this work was undertaken.

\end{document}